\magnification=1200

\pageno=1
\centerline {\bf STRING THEORY, SCALE RELATIVITY AND  }
\centerline {\bf THE GENERALIZED UNCERTAINTY PRINCIPLE    }
\medskip
\centerline {\bf Carlos Castro }
\centerline {\bf Center for Particle Theory }
\centerline {\bf Physics Department}
\centerline {\bf University of Texas}
\centerline {\bf Austin , Texas 78712}
\centerline {\bf ICSC World Laboratory, Lausanne, Switzerland}
\smallskip

\centerline {\bf September, 1995. Revised September 1996 }
\smallskip
\centerline {\bf ABSTRACT}

Extensions (modifications) of the  Heisenberg Uncertainty principle are 
derived 
within the framework of the theory of Special Scale-Relativity  proposed
by Nottale. In particular, generalizations of the Stringy Uncertainty Principle are obtained where the size of the strings is bounded by the Planck scale and the size of the Universe.  Based on the fractal structures inherent with  two dimensional Quantum 
Gravity, which
has attracted considerable interest recently, we conjecture that the underlying
fundamental principle behind String theory should be based on an extension of 
the Scale Relativity principle  where $both$ dynamics as well as scales are incorporated in the
same footing.  

\medskip

\centerline {\bf I. INTRODUCTION}

\medskip
In recent years  considerable attention has been given to the generalizations of Heisenberg's Uncertainty principle and to formulations of nonlinear  Quantum Mechanics [1,2,3]. It was argued  [3] that an extension of quantum mechanics might be required in order to accomodate gravity. Although the current work on string theory suggests that no modifications of Quantum Mechanics might be  necessary to include gravity, evidence of Planck scale physics suggests otherwise.
The poor understanding of physics at very short distances indicates that the small scale structure of space-time might not be adequately described by classical continuum geometry. One new theory consistent with the Standard Model of particle physics is Connes [4] non-commutative 
geometry. Roughly it is  based on the non-commutative algebra of functions defined on a manifold whose 'points' have been smeared ( into operators) according to the spacetime foam picture at tiny distances.   

In the past two years there has been a considerable advance in the non-perturbative behaviour of string theory where  the existence of duality symmetries in the theory  indicate that strings  do not distinguish small spacetime scales from large ones. This is the so-called 
$T$ duality symmetry. This in turn calls for a modification of Heisenberg's Uncertainty principle where beyond Planck scale energies the size of the string grows with momenta instead of the opposite.
For a pedagogical introduction on the new description of spacetime as a result of string theory see Witten [28].

Sometime ago [5,6] argued that some sort of enlarged equivalence
principle is operating in string theory in which dynamics is not only independent
of coordinate transformations but also of structures occurring at distances 
shorter than the fundamental string length, $\lambda_s =\sqrt {2\alpha \hbar}$, in
units $c=1$; i.e. distances smaller than $\lambda_s$ are not relevant in string
theory. Many arguments have been given to support this fact : high energy scattering
at fixed angle [7] ; renormalization group theory analysis based on a discretized version
of Polyakov's generating functional [8,9]; high temperature behaviour of the free
energy [10]; the above-mentioned duality of strings and other higher dimensional extended objects : p-branes or extendons [11]; particle size-growth with momentum
related to information spreading near black hole horizons [12]

This all suggests that below the Planck length the very concept of spacetime changes
meaning and that the Heisenberg Uncertainty principle needs to be modified. The main purpose of this letter is to write down generalizations of the Heisenberg's Uncertainty relations.

One formula
which is a faithful interpolation of the results above is :
$$({\Delta X^\mu \over \lambda_s})\geq({p_s\over \Delta p^\mu}) +({\Delta p^\mu \over
p_s}).~~~p_s\lambda_s =\hbar . \eqno (1)$$

Eq-(1) is valid in flat backgrounds and holds for each spacetime component.

In this letter we will generalize eq-(1) and write down a more accurate equation that encodes the scale relativity principle as well as incorporating the duality principle in string theory.  Despite many efforts,  the fundamental principles underlying string theory are 
still 
unknown, in particular the search for higher symmetries. The  second purpose of this 
letter is to conjecture that the special theory of scale relativity recently proposed
by Nottale [13,14] must play a fundamental role in string theory, specially in
regards to the fact that this theory demonstrates that there is a universal, 
absolute and impassable scale in Nature, which is invariant under dilatations.
This lower limit is the Planck scale.

It was emphasized by Nottale in his book 
that a full motion plus scale relativity including all spacetime components, angles
and rotations remains to be constructed. In particular the general theory of scale
relativity. Our aim is to show that string theory provides an important step 
in that direction and viceversa : the scale relativity principle must be operating
in string theory. The cosmological implications of such principle allowed Nottale
to provide a very simple and elegant proposal for the resolution of the cosmological 
constant problem [14]. In particular, the fundamental scales in Nature are determined
by constraints which are set at $both$ the small and large scales. This is in perfect accordance with the duality principle in string theory.  Applying the
scale relativity principle to the Universe one arrives at the conclusion that there
must exist an absolute, impassable, upper scale in Nature which is invariant 
under dilatations (invariant under the expansion of the Universe) which would 
hold all the
properties of infinity. This upper scale, $L$, defines the "radius" of the
Universe and , when it is seen at its own resolution, it becomes invariant under
dilatations [14]. We must emphasize that scale-relativity does not force the universe to
be closed ; an open expanding universe can have an upper    
characteristic
scale ( invariant under dilatations)  $L$ which holds the property of infinity. 

The term in (1) : $\Delta X^\mu \sim \Delta p^\mu $ is a long distance effect due to the
emergence of classical gravity effects from nonpertubative quantum string theory
dynamics [5,6,7,8,9]. These authors showed that string amplitudes at or above the Planck
energy resummed over all orders of perturbation theory around the flat metric and 
turned out to yield classical gravity effects in appropriate kinematic regions.
The $S$ matrix above Planck energies is dominated by graviton exchange at large
impact parameters. G't Hooft [15] has shown that high energy particle scattering is
dominated by graviton exchange. At energies higher than $m_pc^2$ black hole production
sets in accompanied by coherent emission of real gravitons. In contrast the regime
where $\Delta X^\mu \sim ({ 1/\Delta p^\mu })$ is characterized
by ordinary gauge interactions based on standard Quantum Mechanics. 
So there appears to be two physics regimes which are manifestations of the
same quantum string dynamics. One where energies are much larger than the Planck
energy and other where the energies are much smaller.

Within the context of scale relativity we propose  that the  
$\Delta X^\mu \sim \Delta p^\mu $ 
behaviour originates from the existence of the upper scale in Nature; i.e. it
is also a long distance effect pertaining to classical gravity interactions at 
cosmological
scales; whereas the $\Delta X^\mu \sim {1\over \Delta p^\mu }$ terms are the 
standard Quantum Mechanics results originating from the fractal structure of 
spacetime at microphysical scales [13,14,29]. It is in this fashion how string theory
and scale relativity merge. At first sight it seems surprising how the upper length scale, $L$, 
can have a connection to Planck scale physics. This was the basis of Nottale's proposal to the
resolution of the cosmological constant problem. The relationship to string theory is realized 
via the target space ($T$) duality symmetry : $R\leftrightarrow \alpha'/R$; i.e string theory does not
distinguish one spacetime (large radius) from the other (small radius). 
Fractal structures also 
occur at cosmological scales and for this reason Nottale introduced the upper scale in order to
accomodate the scale relativity principle to the whole universe as well.

The problem of time measurement in quantum gravity has been discussed by [30] and a minimum time interval of the order of the Planck time was found. A review of the series of different physical arguments leading to the Planck scale as the minimum scale in Nature was given by Garay [27]. For a discussion of quantum states for non-perturbative quantum gravity which exhibit a discrete structutr at the Planck scale see Ashtekar et al [27].

The authors in [16,17,18,19] have emphasized that the
fractal structure of spacetime is a crucial question in any theory of Quantum
Gravity. Quantum Mechanics requires a functional average over all possible 
equivalence classes of metrics and it is in this way how the effective dimension
of spacetime is measured. It has been shown  [13] that quantum mechanics "arises" from 
the fractal nature of particle trajectories. In particular, the fractal dimension
of space as well as time turned out to be equal to $2$ for a point 
particle in the relativistic domain. 
Hence, the total fractal space-time dimension is $2+2=4$.  The authors [16] provided
ample evidence that the Hausdorff dimension is $D_H=4$ for two-dimensional quantum
gravity coupled to matter with a central charge $c<1$ whereas the spectral dimension
was $D_S=2$. The fact that the $D_H,D_S$ were constant for any value of 
$-\infty <c<1$  seems to indicate that both dimensions are intrinsic properties
of two-dim quantum gravity independent of the coupling to matter and hence an intrinsic property of the geometry of spacetime.   
The essence of the observations of [13]  was based on the fact that the relevant 
paths in Feynman's path integral description of Quantum Mechanics are the 
fractal trajectories; i.e; continuous but nondifferentiable at any point.

It is the purpose of this letter to bridge both string theory and scale relativity
within the framework of Heisenberg's Uncertainty principle and derive a more
general expression than the one in (1) and the one presented in [5-9].
The expression below Eq-(15) is  very relevant if one were to include the 
fractal nature
of two-dimensional quantum gravity in string theory; i.e. a fractal two dimensional
surface moving in a fractal target spacetime.

The importance of  noncontinuous
maps
in string theory has been discussed in [24]. The space of string configurations
in string theory required both  continuous and noncontinuous square integrable
maps in order to reproduce the results from the dual models.  
The size and shape of strings in their ground state in the lightcone gauge
was investigated in [25]. It was found that in two-dimensions the extrinsic
curvature was divergent. A regularization scheme was used where the string was
kept continuous. As the dimensionality of spacetime increased the string became
smoother and had divergent average size. This is unphysical since their size
cannot exceed the size of the Universe. It is for this reason that the upper scale in nature
must also appear in eq-(15). The average curvature diverges in $D=4$
due to kinks and cusps on a string. It is important to study these properties
further.

It has been suggested by many authors that if one wishes to study Planckian physics 
one must abandon the Archimedean axiom where
any given large segment on a straight line can be surpassed by succesive addition of small segments
along the same line. A non-Archimedean geometry has been proposed using $p-adic$ analysis. For
a review of $p-adic$ strings see [26].  

Having presented the reasons why scale relativity is a relevant issue in string
theory we shall derive the scale-relativity extension of the stringy-uncertainty
principle which is a  more general uncertainty relation than the ones considered so far. 

\smallskip

\centerline {\bf II. The Generalized Uncertainty Principle}
\smallskip
Our goal now is to write down a more precise expression for the enlarged 
uncertainty principle (1) based on the scale relativity principle. The expression below
agrees qualitatively with the results in [20] where for energies not too large 
the size of the interaction region increases linearly with energy while at higher
energies it remains constant. ( see Diagram ).

According to scale relativity all physical quantities depend on the resolution
at which they are observed. At scales of the order $r$ the generalized de Broglie-Compton relation is :

$$ln (m/m_o) ={ln (\lambda_o/r)\over \sqrt 
{1-{ln^2(\lambda_o/r)\over ln^2 (\lambda_o/\Lambda)}}}=ln(\lambda_o/r)^\delta. \eqno (2)$$

The quantities $m_o,\lambda_o $ are the reference mass and  length scales with 
respect to with we measure the other scales. If one chose these scales  to
be the electron's mass and its de Broglie wavelength this would be the signal 
of the quantum mechanical-classical physics
transition of the electron. Scale relativity proposes  that in general $\lambda_o \sim(\hbar/m_o c)$ 
instead of being equal and  the lowest scale is  $\Lambda$, the Planck length 
$\Lambda =\sqrt{(\hbar G/c^3)}=(\hbar/m_pc)$.  The square root factor in the second term of the r.h.s of (2) is a 
Lorentzian-like gamma factor of the same type which appears in special relativity : $m=m_o [{1-v^2/c^2}]^{-1/2}$. Now it   plays in (2)  the 
role of a scale-dependent anomalous dimension $\delta =D_F -D_T$; where 
$D_F$ is the fractal
spatial dimension of a fractal curve trajectory and $D_T$ is the 
topological dimension.
At the quantum-classical transition : $r=\lambda_0$ one has  $D_F=2;D_T=1$ and   $\delta$ becomes equal to $1$ whereas for other values of $r<\lambda_o; \delta>1$. Similar considerations
apply to the time coordinate [13,14].

We can see that if one sets the minimum length to zero, eq-(2) yields the ordinary Compton-de Broglie relation since for $\delta=1$ :

$$  p\lambda=p_o\lambda_o =\hbar                             \eqno (3)$$

Setting $p\sim \Delta p$ and $\delta \sim \Delta x$ in (3) gives the original Heisenberg's uncertainty relation :

$$   \Delta p \Delta x \sim \hbar                         \eqno (4) $$

Whereas from (2) one learns that a  plausible  modification of the Heisenberg''s uncertainty relation is :

$$ln (\Delta p/p_o) ={ln (\lambda_o/\Delta x)\over \sqrt 
{1-{ln^2(\lambda_o/\Delta x)\over ln^2 (\lambda_o/\Lambda)}}}. \eqno (5)$$
Similar considerations apply to the other spacetime coordinates and momenta.

Now we shall concentrate on the large scale behaviour of spacetime. The absolute, impassable, large scale in nature [14] is relevant to fractal structures in cosmology. In particular 
the cosmological generalized-Schwarzschild mass relation is :

$$ln (M/m_g) ={ln (R/\lambda_g)\over \sqrt 
{1-{ln^2(R/\lambda_g)\over ln^2 (L/\lambda_g)}}}. \eqno (6)$$

where $m_g,\lambda_g$ are the mass and length scales which signal the 
classical physics (scale independence) 
cosmological domain transition (where fractal strucutures over large scales  in the Universe become important); the quantity : 
$\lambda_g\sim (Gm_g/c^2)$ where $G$ is Newton's constant and 
 $L$ is the upper, impassable, absolute scale we referred to
earlier which is invariant under dilatations. The value of $L$ was set in [14]
to be $L\sim 10^{61} \Lambda$. To simplify matters we can set $\hbar=c=1$. We also notice
that eqs-(2,6) are valid in any logarithm base, we opt to use the natural base. 

When the upper scale $L\rightarrow \infty$, eq-(6) becomes the original 
Schwarzschild mass relation ( up to a factor of two ) :

$$M\sim {Rc^2\over G}                  \eqno (7)$$

Setting $M\sim \Delta p$ and $R\sim \Delta x$ in (7) one recovers the behaviour of the second term in the r.h.s of (1) signaling the graviton exchange dominance in high energy string scattering . At higher energies than the Planck energy the size of the string $increases$ instead of decreasing. The energy imparted to the string is used to break the strings into pieces. This picture is the one proposed by [12]
consistent with the Bekenstein-Hawking bound for the entropy of a black hole in terms of the horizon's area :
$ S=A/4G      $. This is the statement that one cannot have more than one bit of information per unit area in Planck units.

Inverting the relations given by  eqs-(2,6)
yields :

$$ln (\lambda_o/r) ={ln (m/m_o)\over \sqrt 
{1+{ln^2(m/m_o)\over ln^2 (\lambda_o/\Lambda)}}}. \eqno (8)$$

and :

$$ln (R/\lambda_g) ={ln (M/m_g)\over \sqrt 
{1+{ln^2(M/m_g)\over ln^2 (L/\lambda_g)}}}. \eqno (9)$$

We learnt from (1) that the minimum value of $\Delta X$ is $\sim \lambda_s$.
Lets suppose that  one wishes the two curves given by eqs-(8,9) to intersect at the point 
$(\lambda_p,m_p)$ with $m_p \equiv(1/\Lambda)$,  where 
 $m_o \sim (1/\lambda_o)$ and $\lambda_p \not= \Lambda$!!. 
The latter  occurs because in scale relativity the high energy length and mass scales 
$decouple$ : energy tends to $infinity$ when distances  reach the Planck scale.

The intersection of the two graphs described by (8,9) requires to $extend$ the domain of validity of the scaling exponents appearing in eqs-(2,6) and the 
remaining equations that follow. Therefore we must have the following expressions for the scaling exponents in the extended domains :

$$ \delta_{Compton}   =   {1\over \sqrt 
{1-{ln^2(\lambda_o/r)\over ln^2 (\lambda_o/L)}}}.~r> \lambda_o. \eqno (10)$$
$$ \delta_{Schwarzschild}   =   {1\over \sqrt 
{1-{ln^2(\lambda_g/r)\over ln^2 (\lambda_g/\Lambda)}}}.~r< \lambda_g. \eqno (11)$$

The reason this is necessary is to ensure that the scaling exponents are always real. In ordinary relativity the velocity cannot exceed the speed of light and the Lorentz dilation factor for this reason is never complex.

With this in mind the intersection of the two graphs described by (8,9) in the region $\Lambda<\lambda_p <r<\lambda_o$ requires  
to have :

$$ln (\lambda_o/\lambda_p) ={ln (m_p/m_o)\over \sqrt 
{1+{ln^2(m_p/m_o)\over ln^2 (\lambda_o/\Lambda)}}}. \eqno (12)$$
 
one learns from (12) that $\Lambda <\lambda_p<\lambda_o$. The other relation is

$$ln (\lambda_p/\lambda_g) ={ln (m_p/m_g)\over \sqrt 
{1+{ln^2(m_p/m_g)\over ln^2 (\Lambda/\lambda_g)}}}. \eqno (13)$$

Notice the factor of $\Lambda$ appearing in the scaling exponent of (13) versus the $L$ appearing in 
eq-(9). 
Eliminating $\lambda_p$ from (10,11) gives an algebraic  relationship between the $\lambda_o$
and the $\lambda_g$ scales based on the input  values of $\Lambda,L$. The value of 
$m_g$ is $\sim (\lambda_g/\Lambda^2)$ and of $m_o\sim 1/\lambda_o$. 

We have learnt from scale relativity that the Planck mass $m_p$ does not correspond any more to 
the Planck length $\Lambda$. The Planck length now corresponds to an infinite
mass. In motion special relativity it is required an infinite energy to accelerate
 a particle of non-zero rest mass to the speed of light. Not surprisingly, when we reach the resolution of the Planck's  length an infinite amount of energy
is required.

In order to solve for eqs-(12,13) let us assume that 
$\lambda_s \sim \lambda_p$. 
The string scale $\lambda_s$ is model dependent [8,9]. In the heterotic string
$\Lambda =(\alpha_{GUT}/4)\lambda_s \sim 10^{-2}\lambda_s$. Now let us compute the
fundamental lengths $\lambda_o,\lambda_g$ starting from eqs-(12,13). An estimate is obtained by
setting $ln (\lambda_o/\Lambda)\sim ln (m_p/m_o)$. Eq-(12) becomes in this 
approximation :

$$\lambda_p \sim \lambda_o (\Lambda/\lambda_o)^{1\over \sqrt 2}. \eqno (14)$$

Therefore, if we were to impose  $\lambda_s \sim \lambda_p$,
the length scale $\lambda_o$ is of the order
$\lambda_o \sim 10^7\Lambda \sim 10^{-26}cm$. In any case we have that
$\Lambda<\lambda_s<\lambda_o$. Having fixed $\lambda_o$,  eq-(13) generates the
other scale $\lambda_g$. For this particular choice for $\lambda_p$ one gets 
that $\lambda_g \sim \lambda_o\sim 10^7 \Lambda$, also which would not be very physical.  Clearly our choice of $\lambda_p, m_p$ were  $not$ that appropiate. 

The corect choice of intersection point due to the fact that the scaling exponents are not equal in eqs-(12,13) is to choose
instead $ \lambda_s $  and 
$m_s={\sigma_s\over \Lambda}\not= {1\over \lambda_s}$ 
as the true intersection point. The value of $\sigma_s <1$ so the intersection point lies $below$ the $m_p$ value. Now with the value of 
$\lambda_s \sim 10^{2}\Lambda$ the value of $m_s$ is constrained to obey 
$m_p>m_s>{10^{-2}\over \Lambda}$. Therefore, now 
one would obtain more reasonable values for $\lambda_o,\lambda_g$; ( see diagram ).

In any case one has a free parameter to vary which we can take to be 
$\lambda_s$ assuming the upper scale $L$ is fixed in terms of $\Lambda$. Tuning the value of $\lambda_s$ furnishes a range of values for the transition scales $\lambda_o,\lambda_g$ for $L=10^{60}\Lambda$. Other choices for $L$ leaves us even more freedom to work with.

The diagram referred as figure 1, after the $extension$ of the scaling exponents has been performed in the whole regions of $r$,  
shows the three "quantum", "classical", "cosmological" domains from the smallest 
scale, $\Lambda$ to the largest one $L$. A similar diagram can be found in Nottale's book 
[14 ,ch.7].
The relevance of this diagram is that the region comprised within the two curves
representing the generalized Schwarzschild and Compton-de Broglie  formulae (2,3) is similar
in shape to the allowed region obtained from the Stringy  Uncertainty Principle (1).
This is not to say that both regions are identical !.

Now we are in a position to write down the scale relativity extension/modification  of the Stringy-Uncertainty
Principle. Using eqs-(8,9),   replacing masses for momentum quantities 
, $\Delta p$,  and 
lengths for $\Delta X$ one 
can infer
that :

$${\Delta X^\mu \over \lambda_o }\geq {1\over 2}({p_o\over \Delta p^\mu})^{\alpha} + {1\over 2}
{\lambda_g\over \lambda_o} ({m_o \over m_g})^{\beta} 
({\Delta p^\mu \over p_o})^{\beta}. \eqno (15)$$

This expression is valid for each spacetime component separately. The momentum dependent 
coefficients $\alpha,\beta$ are :

$$\alpha ={1\over \sqrt 
{1+{ln^2(\Delta p^\mu/p_o)\over ln^2 (\lambda_o/\Lambda)}}}.~r<\lambda_o \eqno (16)$$

$$ \beta = {1\over \sqrt 
{1+{ln^2(\Delta p^\mu /m_g)\over ln^2 (L/\lambda_g)}}}.~r>\lambda_g 
\eqno (17)$$. 

and :

$$\alpha ={1\over \sqrt 
{1+{ln^2(\Delta p^\mu/p_o)\over ln^2 (\lambda_o/L)}}}.~r>\lambda_o \eqno (18)$$

$$ \beta = {1\over \sqrt 
{1+{ln^2(\Delta p^\mu /m_g)\over ln^2 (\Lambda/\lambda_g)}}}.~r<\lambda_g 
\eqno (19)$$. 

with the proviso that $\lambda_o,\lambda_g$ are expressed in terms of $\Lambda,L$ as explained earlier.

Lets analyse eq-(15) in detail. Firstly, whenever $\Delta X =\Lambda $ or $L$ the $\Delta p =\infty$ or $0$. The values of $\Delta X$ are confined to the Planck scale and the upper impassible scale whereas there is no ultraviolet nor 
infrared cutoff in the momenta. The limiting behaviour of (15) is :

(i). When $\Delta p^\mu$ is very small the first term in the r.h.s  of (15) dominates and one has 
the standard Heisenberg's relation for $\alpha \rightarrow 1$ when 
$\Lambda \rightarrow 0$. 

(ii). When $\Delta p^\mu$ is very large the second term of
the r.h.s of (15) dominates and one has 
$${\Delta X^\mu \over \lambda_g }\geq({\Delta p^\mu \over m_g c})^{\beta}. \eqno (20)$$
in the $L\rightarrow \infty$;$\beta \rightarrow 1$ and one recovers the regime
where graviton exchange dominates over other interactions and the uncertainty 
grows with momentum : 
$\Delta X\sim \Delta p$.

(iii). The stringy enlarged uncertainty principle is recovered for the following
behaviour in the middle region of the diagram $\lambda_o<r<\lambda_g$ :

$$\alpha,\beta \sim 1.~\lambda_o=a \lambda_s.~ p_o=a^{-1} p_s.~\lambda_op_o=\lambda_s p_s =1~
;
\lambda_o\sim {1\over m_o}.~m_g\sim {\lambda_g\over \Lambda^2};   \eqno (21)$$

so that (15) becomes in that region  :
$${\Delta X^\mu \over \lambda_s }\geq {1\over 2}({p_s\over \Delta p^\mu}) +
{1\over 2} ({\Lambda \over \lambda_s})^2 
({\Delta p^\mu \over p_s}). \eqno (22)$$
with $p_s\lambda_s =1$. The factors of $a$ explicitly cancel.

One recognizes that the "Galilean" limit ( analog of $c\rightarrow \infty$)
 is attained when lowest scale tends to
zero and the upper scale tends to infinity. In eq-(22) we can see that the ratio
$(\Lambda/\lambda_s)^2 $ appearing in the second term of the r.h.s of  (22) tends to one when
the string scale $\lambda_s\rightarrow \Lambda$. Therefore, eq-(22) reduces to eq-(1) in that limit. The results of [5-9] confirm that
the string is not sensitive to scales smaller than $\lambda_s$; i.e scales of the
order of the Planck length. Therefore our results based on scale
relativity  are compatible with the results obtained from string theory.

(iv) A  time-energy uncertainty relation 
 which involves the ratio of two scales 
has been   given by Itzhaki [30] by assuming that the gravitational field fluctuations  due to the mass-energy of a clock ( whose metric  is the Schwarzschild type ) contributes to a time uncertainty of the order  : 
$$\Delta t \ge {\hbar \over \Delta E} +{2Glog (x/x_c)\Delta E\over c^5}. \eqno (23)$$
with $x_c$ being the shortest distance for which general relativity is a good approximation to quantum gravity ( it can be taken as the size of the clock ) and $x$ is the distance from the clock to the observer. Once again, we can see that eq-(19) has the same energy dependence as the second term of eq-(18) and 
of eq-(13) for $\alpha =1;\beta=1$.

To the author one important lesson gained from these results is not only
the role that scale relativity will play in underlying the fundamental
principles to formulate string theory but in the need to modify
Quantum Mechanics in order to accomodate gravity. Some time ago we were able 
 to show how 
Nonlinear Quantum mechanics admitted a geometrical interpretation based on
Weyl geometry [3]. In particular, the nonlinear 
corrections to the Hydrogen atom ground sate energy was $\sim 10^{-35}~ev$.
 For an earlier geometrical origins of Quantum Mechanics based
on Weyl geometry see [22]. Later, Nottale extended the notion of
scale covariance within the fractal nature of spacetime microphysics 
 to account for Quantum
effects. An earlier treatment of fractal trajectories and quantum mechanics was given by G.N. Ord [29].  Since the uncertainty principle has been modified it is not surprising
that one might have to modify accordingly the standard Quantum Mechanics in order to accomodate 
these
results : in particular the standard commutation relations . For a recent proof that $[p,q]\not=ih$ see [23].

Also relevant about the minimum length in Nature are  the arguments of [8,9] pertaining to the nonexistance of 
black holes of sizes $2M$ smaller than $\lambda_s$ and the subsequent fact 
that Hawking radiation must stop at some point when the Hawking temperature reaches
the Hagedorn temperature $T=(1/8\pi M)=T_H$. It was suggested that at this
point conformal invariance is broken and no consistent string propagation is 
possible. 

A more detailed study of fractal surfaces within the context of string theory
is warranted as well as the role that scale relativity will play in a theory
of Quantum Gravity. An extension of scale relativity where both dynamics as 
well as scales are on the same footing should reveal very relevant information
about the underlying principle behind string theory as suggested in [5].
Whether or not Quantum Mechanics remains intact still remains as a challenging question.

\centerline {\bf Acknowledegements}

We are greatfuly indebted to G. Sudarshan and L. Nottale for insights.
Also, to M.Bowers, A.Fuentes, A.Botero and D.Wu for their help . This work was supported in part by a World Laboratory Fellowship,

\centerline {\bf REFERENCES}

1.P. Pfeif, J. Frohlich : Rev. Mod. Physiscs  {\bf 67} no.4 (1995) 759-779

2. S. Weinberg : Annals  Physics {\bf 194} (1989) 336-386

3. C. Castro : Foundations of Physics Letters {\bf 4} no.1 (1991) 81.

4. A. Connes : ``Non-Commutative Geometry'' Academic Press. NY. 1994

5.D. Amati, M. Ciafaloni and G.Veneziano : Phys.Lett. {\bf B 197} (1987) 81.

D. Amati, M. Ciafaloni and G.Veneziano : Phys.Lett. {\bf B 216} (1989) 41.

6.M. Fabrichesi, G.Veneziano : Phys.Lett. {\bf B 233} (1989) 135.

7.D. Gross, P.F. Mende :  Phys.Lett. {\bf B 197} (1987) 129.

8 . K. Konishi, G. Paffuti, P. Provero : Phys.Lett. {\bf B 234} (1990) 276.

9.R. Guida, K. Konishi, P. Provero : Mod. Phys. Lett {\bf A 6} no.16 (1991) 

1487.

10. J.J. Atick, E. Witten : Nucl. Phys. {\bf B 310} (1988) 291.

 K. Kikkawa, M. Yamasaki : Phys.Lett. {\bf B 149} (1984) 357.

N. Sakai, I. Senda : Prog. Theor. Phys. {\bf 75} (1986) 357

12.L. Susskind : " The World as a Hologram " hep-th/9409089.

13.L.Nottale :Int.J.Mod.Phys.{\bf A 4} (1989) 5047.

Int.J.Mod.Phys.{\bf A 7} no.20, (1992) 4899.

14 .L.Nottale: "Fractal Space Time and  Microphysics  : 

Towards

the Theory of Scale Relativity " World Scientific 1992.

15.G't Hooft : Phys.Lett. {\bf B 198} (1987) 61.

16.J. Ambjorn, Y. Watabiki : Nucl. Phys. {\bf B 445} (1995) 129.

17. J. Ambjorn, J. Jurkiewicz, Y. Watabiki : "On the Fractal Structure of 

Two Dimensional Quantum Gravity " hep-lat /9507014. NBI-HE-95-22 preprint.

18. S. Catterall, G. Thorleifsson, M. Bowick, V. John :" Scaling and the 

Fractal Geometry of Two-Dimensional Quantum Gravity " hep-lat/9504009.

19.J. Ambjorn, J. Jurkiewicz : " Scaling in Four Dimensional Quantum Gravity "

hep-th/9503006.

20. P.F. Mende, H. Ooguri : Nucl. Phys. {\bf B 339} (1990) 641.

21.C.Castro : Foundations of Physics vol. {\bf 22} no.4 (1992) 569 

22. E.Santamato : Phys. Rev. {\bf D 29} (1984) 216

23. J. P. Costella : " $[p.q]\not= ih"$  Melbourne preprint. UM-P-95/51.

24. J. Borde, F. Lizzi : Mod. Phys. Lett. {\bf A 5} no.10 (1990) 1911

25. M.Karliner,I.Klebanov, L.Susskind:Int. J. Mod. Phys.{\bf A 3} (1988) 1981.

26. V.S. Vladimorov, I.V. Volovich, E.I. Zelenov : " p-adic Analysis and Mathem

atical Physics "

27. L. Garay : Int. Journ. Mod. Phys {\bf A 10} (1995) 145.

A.Ashtekar, C.Rovelli and L. Smolin : Phys. Rev. Lett {\bf 69} no. 2 (1992)

237.

28. E. Witten : Physics Today,  April 1996, page 24.

29. G.N. Ord : Jour. Physics A. Math General {\bf 16} (1983) 1869.

30. N. Itzhaki : `` Time measurement in Quantum Gravity `` hep-th/9404123.

World Scientific. Singapore 1994.

\end